\documentclass[amsmath,amssymb,aip,jcp,preprint]{revtex4-1}

\usepackage{amsmath,amssymb}
\usepackage{dcolumn}
\usepackage{graphicx}

\begin{document}

\title{Long-range three-body atom-diatom potential for doublet Li${}_3$}

\author{Jason N. Byrd}
\email{byrd@phys.uconn.edu}
\author{H. Harvey Michels}
\author{John A. Montgomery, Jr.}
\author{Robin C\^{o}t\'{e}}
 \affiliation{Department of Physics, University of Connecticut, Storrs, CT 06269}

\begin{abstract}
An accurate long-range {\em ab initio} potential energy surface has been calculated
for the ground state ${}^2A'$ lithium trimer in the frozen diatom approximation using
all electron RCCSD(T).  The {\em ab initio} energies are corrected for
basis set superposition error and extrapolated to the complete basis
limit.  Molecular van der Waals dispersion coefficients and three-body
dispersion damping terms for the atom-diatomic dissociation limit are presented
from a linear least squares fit and shown to be an essentially exact
representation of the {\em ab initio} surface at large range.
\end{abstract}

\maketitle

\section{Introduction}

Progress in the field of ultracold molecules has been rapidly growing for the
past decade, as many atomic and molecular research groups have turned towards
the study of either production or dynamics of ultracold diatomic molecules.  The
success behind the study of ultracold molecular formation lies in the use of
photoassociation and Feshbach resonances (see Jones {\em et al.}
\cite{jonesphoto} and K{\"o}hler {\em et al.} \cite{kohlerfesch} for recent
reviews).  Using a combination of photoassociation and STIRAP\cite{stirap}
(stimulated Raman adiabatic passage) the formation of vibrational ground state
KRb\cite{ospaelkaus2008,ni2008} and Cs${}_2$\cite{gustavsson} molecules have
been reported.  With this prospect of ground vibrational diatoms in mind we
focus our study to the interaction effects of a colliding lithium atom with that
of a $v=0$ singlet lithium diatom.  Knowledge of both the long- and
short-range\cite{byrd2009-a} interaction potential can be used to calculate
inelastic three body rates\cite{cvitas2007} as well as investigate Efimov
collisions.\cite{nielsen2002}

The assumption of a vibrational ground state diatom greatly simplifies the physics of the
interaction, however the complexity of calculating long-range interactions for even
the $v=0$ range of motion of the diatom is significant.  
Given this we work within the frozen diatom approximation, which entails
freezing the diatom at the calculated equilibrium bond length for the entire
potential energy surface, which is analogous to vibrationally averaging the
diatom over the course of the collision.  The long-range interaction is not strongly
effected by this approximation due to the small linear changes seen in the diatomic
polarizability as the diatom undergoes small oscillations. Here then the
polarizability averaged over the vibrational motion is just the polarizability
evaluated at $r_e$.  The validity of the rigid rotor approximation for short
atom-diatom collisional distances was found to be good for distances greater
than $10$\AA.

We structure this work into the following three parts, first is the discussion
of the {\em ab initio} calculations.  Here we discuss the methodology involved
in choosing the appropriate basis set which provides both the optimal values for
the spectroscopic constants as well as the best atomic static polarizability.
Further improvement to the interaction energy is shown to come from accounting
for basis set superposition error through a counterpoise correction\cite{bsse}
and from extrapolating the counterpoise corrected energies to the complete
basis limit.  Next we present a summary of the long-range van der Waals
interaction for the tri-atomic dissociation limit.  An account of the many-body
terms arising from third and fourth order Rayleigh-Schr\"{o}dinger perturbation
%theory that contribute to the three-atom $C_6$, $C_8$ and $C_{10}$ van der Waals
theory that contribute to the three-atom $C_6$ and $C_8$ van der Waals
coefficients is given.  Finally we present the long-range analytic van der Waals atom-diatom
interaction energy of Cvita\v{s} {\em et al.}\cite{cvitas2} and 
our fitted non-additive van der Waals coefficients.

\section{{\em Ab initio} Calculation}

The long-range potential energy surface for the ${}^2A'$ Li${}_3$ state has been
calculated within a frozen diatom approximation for collisional angles near the
$C_{2v}$ geometry, which corresponds to a lithium diatom in the
$X{}^1\Sigma^{+}_{g}$ state colliding with a single ${}^2S$ state lithium atom.
The long-range potential energy surface was calculated for atom-diatom distances
ranging from $10$\AA~to $100$\AA, a region we consider to be both outside of any
consideration of charge overlap yet still well within the very-long-range limit
where the retarded potential starts contributing.  The collisional angle
sampling of $\theta=60,~70,~80\text{ and } 90$ degrees (where
$\theta=90^{\circ}$ corresponds to $C_{2v}$ geometry in Jacobi $R,~r,~\theta$
coordinates) is both sufficient for an accurate fit and consistent with our
previous work on the near equilibrium geometry potential energy
surface.\cite{byrd2009-a}  All electronic energy calculations in this work were
done correlating all electrons using spin restricted coupled cluster theory with
singles, doubles and iterative triples\cite{knowles1993,knowles2000} (RCCSD(T))
as implemented in the MOLPRO 2008.1\cite{molpro08short} suite of {\em ab initio}
programs.

Cold scattering calculations require exceptionally accurate interaction
potentials in order to properly predict cross-sections and scattering lengths.
To provide this accuracy we apply a series of corrections to the RCCSD(T) energy
which account for deficiencies within the basis set.  Additionally we correlate
all electrons in the RCCSD(T) energy so as to account for the core-core and
core-valence (CV) contributions.  The inclusion of CV correlation energy has
been shown to account for roughly $0.002$\AA~for multiple bonds and several
hundred cm${}^{-1}$ to atomization energies.\cite{martin1995}  To properly correlate
all electrons within an {\em ab initio} calculation necessitates the use of a CV
consistent basis
set.\cite{martin1995,iron2003}  We have examined the use of the four and five zeta
cc-pVnZ basis sets from Feller\cite{feller} and the CV consistent CVnZ basis sets
of Iron {\em et al.}\cite{iron2003}  In Table \ref{diaspectro} is a comparison
between calculated spectroscopic constants from the above basis sets and the
experimental values, it can be seen that the five zeta basis sets provide a
marked improvement in terms of both $r_e$ and $D_e$.  
The calculated polarizability is also used as a benchmark calculation in
addition to $r_e$ and $D_e$.  The accuracy of which is a strong indicator of the
accuracy of the atomic static polarizability is a strong indicator of the
accuracy of long-range molecular dispersion interactions.

The first improvement to the trimer potential energy surface was to calculate
the basis set superposition error (BSSE) through a counterpoise
calculation.\cite{bsse}  The interaction energy of the trimer is then
\begin{equation}
E_{\rm int} = E_{\rm trimer}-E^{12}_{\rm atom}-E^{13}_{\rm
	atom}-E^{23}_{\rm atom}
\end{equation}
where $E^{ij}_{atom}$ is the energy of the resulting atom when atoms $i$ and
$j$ within the trimer are replaced with dummy centers.  Accounting for BSSE the CVQZ
basis provides a correction of $5.27$ cm${}^{-1}$ and the CV5Z basis has a
correction of $1.05$ cm${}^{-1}$.  Across the total long-range surface the total
BSSE correction varies no more than 1\%, finally converging to a constant value
at interaction regions greater than $20$\AA.
This suggests that the majority of the BSSE
corresponds to the diatom and not the atom-diatom interaction itself.  
Still this counterpoise correction
does not fully account for all of the dissociation energy of the Li${}_2$
diatom.  We further improve upon the accuracy of the potential energy surface by
extrapolating the counterpoise corrected energies to the complete basis set
(CBS) limit.  We use the CBS limit extrapolation formulation of
Helgaker {\em et al.}\cite{helgaker1997}
\begin{equation}\label{cbseqn}
E_n=E_{\infty} + \frac{a}{n^3}.
\end{equation}
This extrapolation scheme was applied to the CVQZ and CV5Z counterpoise corrected
interaction energies for the final potential energy surface calculation.  

We calculate the static polarizability for the ground state lithium atom and singlet
lithium diatom with the static field method\cite{LFK1,LFK2} by calculating the RCCSD(T)
energies in the presence of an electric dipole field (given here as $E(F)$).  The static
polarizability is given by the finite field gradient
\begin{equation}
\alpha_{ii} = \frac{5E(0)/2 - 8E(F)/3 + E(2F)/6 }{F^2}
\end{equation} 
and reported in Table \ref{polarizability}.  The dispersion energy between two
monomers at long-range (no charge overlap) can be expressed
as\cite{buckingham1967}
\begin{equation}
u_{disp}\simeq\frac{3 V\alpha_1\alpha_2}{2R^{6}},
\end{equation}
where $\alpha_i$ is the static polarizability for the $i$th monomer and $V$ is a
characteristic excitation energy of the molecule\footnote{The characteristic
energy is typically evaluated using Uns$\rm \ddot{o}$ld's approximation as $V=U_1
U_2/(U_1+U_2)$ where $U_i$ is usually set to the first two excitation 
energies.\cite{buckingham1967}}.  Due to the important contribution of the long-range tail
of the electron wave function to the molecular polarizability, the effects of
adding diffuse functions can be significant.  In the calculation of the
polarizability discussed above, a set of even tempered diffuse functions were
added to the CVnZ basis sets and found to contribute little to the extrapolated
polarizability.  A further test on the effect of adding diffuse functions was performed
by evaluating a CBS extrapolated single point energy calculation at $10$\AA~with
the even tempered diffuse functions discussed above.  The results showed that in
the CBS limit the difference between the standard and augmented CVnZ basis sets
amounts to less than half a wavenumber.

Because the long-range interaction depends explicitly upon the monomer static
polarizability, it is clear from the reported atomic static polarizability in
Table \ref{polarizability} that the CVnZ basis sets are the optimal.  The
variation in theoretically reported parallel and perpendicular polarizabilities
for the lithium diatom (see Deiglmayr {\em et al}\cite{deiglmayr08} for a review
of alkali diatomic static polarizabilities) do not offer a significant constraint on
the chosen basis set.  Our final recommended calculations make use of the
counterpoise correction with the extrapolation to the CBS limit as
discussed above.  We expect that our calculated diatomic long-range interactions
will be suitably accurate given the precision of the calculated spectroscopic
values and static atomic polarizability.

\section{long-range van der Waals Potential}

We now examine the trimer potential energy surface at the three atom
dissociation limit, which is expanded analytically in terms of the two-body
dispersion interactions with the addition of a purely three-body interaction
potential.  The three-body contribution to the interaction energy is well known
to be strong for Li${}_3$, for both the doublet\cite{byrd2009-a} and
quartet\cite{pack1} state.  Thus it is important to accurately include such
effects in any model of the long-range interaction.  By using perturbation
theory to examine this expansion it is possible to express the generalized
dispersion coefficients in terms of known diatom and triatomic constants.  Our
goal in this section is to overview the essential theory of three-body atomic
interactions, which can then be specialized to the case of atom-diatom
interactions within the frozen diatom approximation previously discussed.

The tri-atomic dissociation interaction potential can be described in terms of
the diatomic van der Waals interaction potential $V_d({\bf r}_{ij})$ and
non-additive many-body potential $V_3({\bf r})$ as
\begin{equation}\label{thestart} V({\bf r}) = \sum_{i>j} V_d ({\bf r}_{ij}) +
V_3({\bf r}), 
\end{equation} where ${\bf r}=({\bf r}_{12},{\bf r}_{13},{\bf
r}_{23})$ are the three internuclear vectors.  Long-range dispersion interaction
potentials (excluding retardation and orbital overlap effects) between two
S-state atoms are described using the multipole expansion
\begin{equation}\label{vdw} V_d(r_{ij})=-C_6 r^{-6}_{ij}-C_{8}
r^{-8}_{ij}+O(r^{-10}_{ij}), 
%r^{-8}_{ij}-C_{10} r^{10}_{ij}+O(r^{-12}_{ij}), 
\end{equation} 
where the $C_6$ and $C_8$ coefficients are respectively the dipole-dipole,
quadrupole-quadrupole and dipole-octopole expansions\cite{buckingham1967} of the
inter-atomic electrostatic Hamiltonian $V_{ij}$.  To obtain the leading terms of
the non-additive potential $V_3({\bf r})$ in Eq. \ref{thestart}, the analogous
%$C_8$ and $C_{10}$ coefficients are respectively the dipole, quadrupole and
%octopole dipolar expansions\cite{buckingham1967} of the inter-atomic
%electrostatic Hamiltonian $V_{ij}$.  To obtain the leading terms of the
expansion method used for the diatom van der Waals interaction can be used.
Here, Rayleigh-Schr\"{o}dinger (RS) perturbation theory is applied to the total
inter-atomic interaction Hamiltonian $H_{\rm int}=H_{\rm tot}-H_A-H_B-H_C$, where $H_x$ is the
atomic Hamiltonian for the $x$'th atomic, then expanded by multipole moments.
The desired contributions to the non-additive potential $V_3({\bf r})$  arise
from the first many-body terms in third order RS perturbation theory.

The bipolar expansion of the many-body terms from third order RS perturbation
theory leads to a summation of purely geometric factors with 
interaction constants dependent on the atomic species\cite{bell,doran}, 
\begin{equation}\label{triplequicky}
V^{(3)}_3 = \sum_{l_1 l_2 l_3} W_{l_1 l_2 l_3}({\bf r}) Z_{l_1 l_2 l_3}.
\end{equation}
The interaction constant can be expressed with the Casimir-Polder integral 
of the dynamic $2^{l_i}$ polarizabilities\cite{mclachlan}
over complex frequencies 
\begin{equation}
Z_{l_1 l_2 l_3} = \frac{1}{\pi}\int_{0}^{\infty}
\alpha^{l_1}(i\omega)
\alpha^{l_2}(i\omega)
\alpha^{l_3}(i\omega)d\omega.
\end{equation}
With the dynamic polarizability defined as\cite{dalgarno1996}
\begin{equation}
\alpha^{l}(\omega)=\sum_n \frac{f^{(l)}_{n0}}{E^2_{n}-\omega^2}
\end{equation}
and the $2^l$ oscillator strength defined as
\begin{equation}
f^{(l)}_{n0}=\frac{8\pi}{2l+1}E_{n}
|\langle 0|\sum_i r^{l}_{i}Y_{lm}(\hat{\bf r_i})|n\rangle|^2,
\end{equation}
the sum over $i$ goes over all the electrons in the given atom and $E_n$ is
again the excitation energy for state n.  The geometric factors $ W_{l_1 l_2
l_3}({\bf r})$ have been reported by a number of
authors\cite{bell,mclachlan,doran,cvitas2} and will not be reproduced here.  The
first term in the third order expansion can be identified as the well known
Axilrod-Teller-Muto triple-dipole\cite{atm} term.  Additionally, it has been
identified that the quadruple dipole term $Z_{1111}$, from fourth order
perturbation theory, has a contribution to the van der Waals dispersion
coefficients of consideration here.  This term has no exact expression in terms
of the dynamic polarizabilities\cite{cvitas2}, but it can be approximated using
a Drude oscillator model in the case of three $S$ state atoms.  Using the
corresponding Drude oscillator approximation for the $C_6$ van der Waals
dispersion coefficient, the $Z_{1111}$ term can be approximated as\cite{bade}
\begin{equation}
Z_{1111}=10Z_{111}^{2}/(3 C_6).
\end{equation}
Using the values $C_6=1393.39$ and $Z_{111}=v_{abc}/3=56865$ from Yan {\em et
al.}\cite{dalgarno1996}, this provides the approximate value $Z_{1111} =
7735638$ a.u. for the quadrupole dipole term.

In the other asymptotic limit of the trimer where the system dissociates to a
diatom and separated atom, the van der Waals type interaction can again be expressed
as a series of multipole terms of the diatomic and atomic polarizabilities.  Using the
Jacobi coordinates $R,~r~\text{and}~\theta$ to describe the atom-diatom system, where
$r$ is the diatomic internuclear distance, $R$ is the diatomic center of mass to
colliding atom distance and $\theta$ is the angle between $r$ and $R$, for the
asymptotic limit of $R\gg r$ the interaction potential in the absence of
damping and exchange is\cite{buckingham1967}
\begin{equation}\label{dalong}
V(R,r,\theta)=D_e-C_{6}(r,\theta)R^{-6}-C_{8}(r,\theta)R^{-8}
+O(R^{-10})
%-C_{10}(r,\theta)R^{-10}+O(R^{-12})
\end{equation}
Where $D_e$ is the interaction energy of the diatom, and the dispersion coefficients
are defined in terms of Legendre polynomials as
\begin{flalign}
\label{general6}
C_{6}(r,\theta) =& C^{0}_{6}(r)+C^{2}_{6}(r)P_2(\cos\theta),\\
\label{general8}
C_{8}(r,\theta) =& C^{0}_{8}(r)+C^{2}_{8}(r)P_2(\cos\theta) +
C^{4}_{8}(r)P_4(\cos\theta)\\
%C_{10}(r,\theta) =& C^{0}_{10}(r) + C^{2}_{10}P_2(\cos\theta)+\nonumber\\
%\label{general10}
%&C^{4}_{10}(r)P_4(\cos\theta) + C^{6}_{10}(r)P_6(\cos\theta).
\end{flalign}
To obtain the analytic form for the atom-diatom van der Waals
coefficients in terms of the tri-atomic terms, we transform the
internuclear $r_{ij}$ coordinates to Jacobi coordinates through the
transformations
\begin{align}
r_{12}&= r,\\
r_{23}&= (R^2+\frac{r^2}{4}+ Rr\cos\theta)^{1/2},\\
r_{31}&= (R^2+\frac{r^2}{4}- Rr\cos\theta)^{1/2}
\end{align}
in the limit of $R\gg r$.  

The contributions to $C_{6}(r,\theta)$ and $C_{8}(r,\theta)$ can be found by
transforming the $W_{l_1 l_2 l_3}({\bf r})$ and $r_{ij}^{-n}$ coefficients
and then
expanding in a power series with respect to $r/R$.  This expansion has been
completed by Cvita\v{s} {\em et al.}\cite{cvitas2} for the contributions to
$C_{6}(r,\theta)$ and $C_{8}(r,\theta)$.  The resulting terms included in these
van der Waals coefficients are
\begin{flalign}
\label{c06}
C^{0}_{6}(r)=& 2 C_6 + 12 Z_{1111} r^{-6} + O(r^{-8}),\\
\label{c26}
C^{2}_{6}(r)=& 6 Z_{111} r^{-3} + 6 Z_{1111} r^{-6} + O(r^{-8}),\\
C^{0}_{8}(r)=& 2 C_8 + \frac{5}{2} C_6 r^2 - \frac{3}{2} Z_{111} r^{-1}
+\nonumber\\& 
\label{c08}\frac{33}{2} Z_{1111} r^{-4} + O(r^{-6}),\\
C^{2}_{8}(r)=& 8 C_6 r^{2} + \frac{6}{7} Z_{111} r^{-1} + \frac{120}{7} Z_{112}
r^{-3} + \nonumber\\& 
\label{c28}
\frac{402}{7} Z_{1111} r^{-4} + O(r^{-6}) \\
\intertext{and}
C^{4}_{8}(r)=& \frac{36}{7} Z_{111} r^{-1} - \frac{120}{7} Z_{112} r^{-3} +
\frac{144}{7} Z_{1111} r^{-4} + \nonumber\\ 
\label{c48}
& (40 Z_{113} - 30 Z_{122}) r^{-5} + O(r^{-6}).
\end{flalign}

In the atom-diatom dissociation limit the effects of charge overlap damping in
the $r$
dependent terms of Eqs. \ref{c06}-\ref{c48} cannot be neglected.  The
standard method of accounting for charge overlap is the inclusion of a damping
coefficient $f_n(r_{ij})$ for each $r_{ij}^{-n}$ contribution.  Recent
reports\cite{rerat2003,cvitas2} have made use of the popular Tang and Toennies\cite{tang}
dispersion damping functions
\begin{equation}
f_n(r_{ij}) = 1-e^{-b r_{ij}}\sum_{k=0}^{n}\frac{(br_{ij})^k}{k!}.
\end{equation}
In this work we chose to follow R{\'e}rat and Bussery-Honvault\cite{rerat2003}'s implementation of
the Tang and Toennies damping functions, where each anisotropic contribution in terms of $r$ has the
following mapping
\begin{equation}\label{tt}
r^{-n}\rightarrow f_n(r)r^{-n}.
\end{equation}
Working within the frozen diatom approximation the charge overlap damping can be
simply modeled as a constant fitting parameter, $F_n$, evaluated
at the diatomic equilibrium bond length.  Inserting the fitting parameter into
Eqs. \ref{c06}-\ref{c48} provides the following 
\begin{flalign}
\label{fc06}
C^{0}_{6}(r_e)=& 2 C_6 + 12 Z_{1111} F_6 r_e^{-6} + O(r^{-8}),\\
\label{fc26}
C^{2}_{6}(r_e)=& 6 Z_{111} F_3 r_e^{-3} + 6 Z_{1111} F_6 r_e^{-6} + O(r^{-8}),\\
C^{0}_{8}(r_e)=& 2 C_8 + \frac{5}{2} C_6 F_2 r_e^2 - \frac{3}{2} Z_{111} F_1 r_e^{-1}
+\nonumber\\& 
\label{fc08}\frac{33}{2} Z_{1111} F_4 r_e^{-4} + O(r^{-6}),\\
C^{2}_{8}(r_e)=& 8 C_6 F_2 r_e^{2} + \frac{6}{7} Z_{111} F_1 r_e^{-1} + \frac{120}{7} Z_{112}
F_3 r_e^{-3} + \nonumber\\& 
\label{fc28}
\frac{402}{7} Z_{1111} F_4  r_e^{-4} + O(r^{-6}) \\
\intertext{and}
C^{4}_{8}(r_e)=& \frac{36}{7} Z_{111} F_1 r_e^{-1} - \frac{120}{7} Z_{112} F_3 r_e^{-3} +
\frac{144}{7} Z_{1111} F_4 r_e^{-4} + \nonumber\\ 
\label{fc48}
& (40 Z_{113} - 30 Z_{122}) F_5 r_e^{-5} + O(r^{-6}).
\end{flalign}
As evaluated in Eqs. \ref{fc06}-\ref{fc48}, the fitting parameter $F_n$ no
longer directly correspond to the Tang and Toennies damping function as has been
noted.\cite{cvitas2}  

\begin{figure}[t]
\resizebox{8cm}{!}{\includegraphics{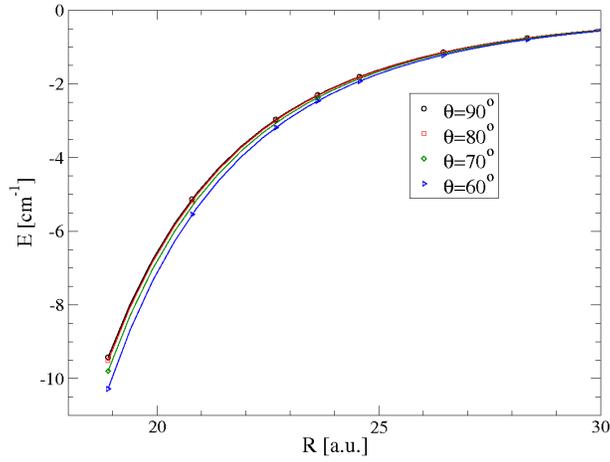}}
\caption{\label{afit}Comparison of the {\it ab initio} surface (points) and
fitted van der Waals potential (solid lines) for the Li [${}^2$S]$+$ Li${}_2$
[${X}^1\Sigma^{+}_{g}$] interaction.  The Li${}_2$ bond length is held at the
calculated equilibrium bond length $r_e$.}
\end{figure}

We have performed a linear least squares fit of Eq. \ref{dalong}, with the
definitions given by Eqs. \ref{general6}-\ref{general8}, to obtain the van der
Waals coefficients given in Table \ref{coefs}.  We have also fitted Eq.
\ref{dalong} with the definitions given in Eqs. \ref{c06}-\ref{c48}.  
We use the diatom $C_6=1393.39$ and
$C_8=8342$ dispersion coefficients as well as the $Z_{111}=v_{abc}/3=56865$
triple dipole term from Yan {\em et al.}\cite{dalgarno1996} and the
$Z_{112}=581000$, $Z_{113}=1.7\times10^7$ and $Z_{122}=6.41\times 10^6$ from
Patil and Tang.\cite{patil}   Fitting to these values we obtain the same
dispersion coefficients as above due to the nature of the least squares fit,
with the values for the damping parameters given in Table \ref{coefs}.  Plotted
in Fig. \ref{afit} is a comparison of the {\it ab initio} surface and the fitted
van der Waals potential.  As can be seen the fit is very accurate, which is
confirmed by the RMS surface fitting error of $10^{-3}$ cm${}^{-1}$.  This
analytical long-range expansion can be used in conjunction with our previous
work\cite{byrd2009-a} on the ground state surface of ${}^2A'$ Li${}_3$ to
evaluate scattering properties of the Li$+$Li${}_2$ rigid rotor system.  The
results of which we leave to future work.

\section{Conclusions}

We have calculated {\it ab initio} an accurate long-range ${}^2A'$ Li${}_3$
surface for the dissociation to the Li [${}^2$S]$+$ Li${}_2$
[${X}^1\Sigma^{+}_{g}$].  The lithium diatom was taken to be rigid rotor with
the bond length constrained to $r=r_e$, the calculated equilibrium bond length
in table \ref{diaspectro}.  The surface was calculated at the RCCSD(T) level of
theory, correlating all electrons.  To accurately describe the CV interaction,
the CVQZ and CV5Z basis sets of Iron {\it et al.} \cite{iron2003} were used; the
final counterpoise corrected interaction energies were then extrapolated to the
CBS limit.  At this level of theory, the atomic and diatomic dipole
polarizabilities are shown to be in good agreement with published experimental
and theoretical results, which is an important component to long-range
interactions.  Using the expansion for the atom-diatom many body van der Waals
potential (Eq. \ref{dalong} and \ref{general6}) we calculate the non-additive
interaction coefficients by fitting to the calculated {\it ab initio} surface.
The resulting repulsive contribution of the three-body interaction to the total
interaction energy is found to be up to 33\% of the total energy.  The fitted
van der Waals coefficients were found to be consistent with the existing
literature on the related ${}^4A'$ state.\cite{rerat2003,cvitas2}

\section{Acknowledgments}

This work was funded in part by the U.S. Department of Energy
Office of Basic Energy Sciences.

%\bibliographystyle{ijqc}
%\bibliography{library}

\newpage

\begin{table}[t]
\begin{tabular}{c|lll}
\hline
  basis
  & \multicolumn{1}{c}{$r_e~$ \AA} 
  & \multicolumn{1}{c}{$D_0~$ cm${}^{-1}$}
  & \multicolumn{1}{c}{$D_e~$ cm${}^{-1}$}\\
 \hline
Exp. 
& 2.673\footnote{reference \cite{huberherzberg}}
& 8434.58\footnotemark[1]
& 8516.78\footnote{reference \cite{robinrkr}} 
\\
\hline
Extrapolated & 2.674 & & 8510.926 \\
CV5Z & 2.674 & 8310.85 & 8487.402 \\
CVQZ & 2.676 & 8288.66 & 8465.005 \\
cc-pV5Z & 2.680 & 8319.12 & 8470.085 \\
cc-pVqZ & 2.685 & 8270.86 & 8445.741 
\\\hline
\end{tabular} 
\caption{\label{diaspectro} Spectroscopic constants for the ground state singlet
lithium diatom calculated using four and five zeta correlation consistent basis sets
with counterpoise.}
\end{table}

\newpage

\begin{table}[t]
\begin{tabular}{cllll}
\hline
  & \multicolumn{1}{c}{Li [${}^2$S]}
  & \multicolumn{2}{c}{Li${}_2$ [${X}^1\Sigma^{+}_{g}$]}  & \\
  basis
  & \multicolumn{1}{c}{$\alpha~$a${}^3_0$}
  & \multicolumn{1}{c}{$\alpha_{\perp}~$a${}^3_0$}
  & \multicolumn{1}{c}{$\alpha_{\parallel}~$a${}^3_0$}\\
 \hline
Extrapolated & 164.1 & 165.6 & 296.1 \\
CV5Z & 164.2 & 165.1 & 296.7 \\
CVQZ & 164.3 & 164.6 & 297.2 \\
cc-pV5Z & 165.6 & 166.4 & 299.6 \\
cc-pVqZ & 166.7 & 166.4 & 300.8 \\
Theory & 164.111\footnote{reference \cite{dalgarno1996}} & & \\
& 164.4\footnote{reference \cite{deiglmayr08}} 
& 162.4\footnotemark[1] & 305.2\footnotemark[1]\\
Experiment & 164.2(1.0)\footnote{reference \cite{miffre}} & & 
\\\hline
\end{tabular}
\caption{\label{polarizability} Calculated finite field static polarizabilities for
the ground state lithium atom and singlet lithium diatom evaluated at the
RCCSD(T)/CV5Z equilibrium bond length.}
\end{table}

\newpage

\begin{table}[t]
\begin{tabular}{cl|cl}
\hline
%$C^{0}_{6}(r_e)=$ & 1696       & $F_1(r_e)=$    & 0.8357\\
%$C^{2}_{6}(r_e)=$ & 278        & $F_{-2}(r_e)=$ & -0.0329\\
%$C^{0}_{8}(r_e)=$ & 143063     & $F_3(r_e)=$    &  0.0462\\
%$C^{2}_{8}(r_e)=$ & 49478      & $F_4(r_e)=$ & 0.0015\\
%$C^{4}_{8}(r_e)=$ & 39942      & $F_5(r_e)=$ &-0.0090\\
%$C^{0}_{10}(r_e)=$ & 9055101   & $F_6(r_e)=$ &-0.0043\\
%$C^{2}_{10}(r_e)=$ & -2243504  & &\\
%$C^{4}_{10}(r_e)=$ & -26042500 & &\\
%$C^{6}_{10}(r_e)=$ & -6718528  & &
$C^{0}_{6}(r_e)=$ & 1604       & $F_1=$    & -0.2823 \\
$C^{2}_{6}(r_e)=$ & 120        & $F_2=$ & 0.1602  \\
$C^{0}_{8}(r_e)=$ & 221646     & $F_3=$    & 0.0399  \\
$C^{2}_{8}(r_e)=$ & 173870     & $F_4=$    & 0.0167  \\
$C^{4}_{8}(r_e)=$ & 27104      & $F_5=$    & 0.0075  \\
                  &            & $F_6=$    & -0.0047 \\
\\\hline
\end{tabular}
\caption{\label{coefs}
Fitted van der Waals coefficients in a.u. for the Li$+$Li${}_2$ ridged rotor system.}
\end{table}

\newpage

\end{document}